\documentclass[10pt, conference]{IEEEtran}
\usepackage{cite}
\usepackage{amsmath,amssymb,amsfonts}
\usepackage{bbm}
\usepackage{algorithmic}
\usepackage{graphicx}
\usepackage{textcomp}
\usepackage{xcolor}
\usepackage{booktabs}
\usepackage{verbatim}
\usepackage{float}
\usepackage{multirow}
\usepackage{hyperref}
\usepackage{setspace}

\begin{document}

\title{A Time-Frequency Generative Adversarial based method for Audio Packet Loss Concealment}

\author{\IEEEauthorblockN{\textit{Carlo Aironi},
\textit{Samuele Cornell},
\textit{Luca Serafini},
\textit{Stefano Squartini}}
\IEEEauthorblockA{Department of Information Engineering,
Università Politecnica delle Marche, Italy\\
Email: (c.aironi, s.cornell)@pm.univpm.it, l.serafini@staff.univpm.it,
s.squartini@univpm.it}}

\maketitle

\begin{abstract}
Packet loss is a major cause of voice quality degradation in VoIP transmissions with serious impact on intelligibility and user experience. This paper describes a system based on a generative adversarial approach, which aims to repair the lost fragments during the transmission of audio streams.
Inspired by the powerful image-to-image translation capability of Generative Adversarial Networks (GANs), we propose \emph{bin2bin}, an improved pix2pix framework to achieve the translation task from magnitude spectrograms of audio frames with lost packets, to non-corrupted speech spectrograms. In order to better maintain the structural information after spectrogram translation, this paper introduces the combination of two STFT-based loss functions, mixed with the traditional GAN objective. 
Furthermore, we employ a modified PatchGAN structure as discriminator and we lower the concealment time by a proper initialization of the phase reconstruction algorithm.
Experimental results show that the proposed method has obvious advantages when compared with the current state-of-the-art methods, as it can better handle both high packet loss rates and large gaps. We make our code publicly available at: \texttt{github.com/aircarlo/bin2bin-GAN-PLC}.
\end{abstract}

\section{Introduction}\label{sec:introduction}

Speech signals are often subject to localized distortions or even total loss of information, when data is transmitted through unreliable channels. This happens, for example, in  applications such as mobile digital communications, videoconferencing systems and Voice over Internet Protocol (VoIP) calls. In such scenarios, audio frames are often encapsulated into packets, which are then routed individually through the network, sometimes taking different paths, resulting in out-of-order delivery. At the destination, the original sequence may be reassembled in the correct order, based on the packet sequence numbers. Hence, a variety of issues can occur, like packet losses, over-delay or jitter.

The process of restoration of missing packets is known as Packet Loss Concealment (PLC)\cite{mohamed2020concealnet}. This term refers to any technique that attempts to overcome the packet-loss problem, by concealing the lost fragments by an estimated reconstruction, which should be meaningful and consistent with the informative content of the speech message. The system should also prevent audible artifacts and decrease listening fatigue, so that the listener remains unaware of any problems that have occurred.


\subsection{Related works}
Some techniques refer to a similar task with the terms Audio Inpainting \cite{adler2011audio, mokry2020audio}, Waveform Interpolation \cite{lagrange2005long} or Extrapolation \cite{maher1994method}. These techniques address the reconstruction problem from a sparsity point of view, by approximating the waveform with a combination of frequency atoms, extracted from a given dictionary. However they are not suitable for real-time applications, as the computational cost can lead to excessive latency times.

Most of the current approaches to PLC are based on codecs that implement algorithmic solutions: sender-based techniques like Interleaving and Forward-Error Correction (FEC) \cite{chua2006qos}, or receiver-based concealment techniques, like Silence/Noise Substitution, Waveform Substitution, or Linear Predictive Coding (LPC) \cite{marafioti2019context}.


With the rise of Deep Neural Networks (DNN), a significant improvement of quality has been obtained on speech processing tasks, hence also DNN architectures for neural PLC have been successfully investigated: MLP \cite{lee2015packet}, LSTM/RNN \cite{lotfidereshgi2018speech, mohamed2020concealnet}, Autoencoders \cite{chang2019deep, kegler2019deep}, GANs \cite{wang2021temporal, ebner2020audio}.

In this study, we apply a Generative method, based on the pix2pix \cite{isola2017image} framework which exploits a Fully Convolutional Network (FCN) architecture, to address the spectrogram inpainting task. We show that this solution, while preserving global temporal and spectral information along with local information, can outperform competing approaches, based either on classical digital signal processing solutions or learning methods.

\section{Generative Adversarial Networks}\label{sec:gans}
Generative Adversarial Networks (GANs) \cite{goodfellow2020generative} have emerged in the past years as a powerful generative modeling technique. A typical GAN consists of two networks, a generator (\emph{G}) and a discriminator (\emph{D}). Given an input of random values sampled from a normal distribution, \emph{z} (latent variable), the generator performs an upsampling in order to obtain a sample of suitable dimensions. On the other hand, the discriminator acts as a binary classifier, trying to distinguish ``real'' samples $x$ (belonging to the dataset distribution) from ``fake'' samples, generated by \emph{G}.

Both \emph{G} and \emph{D} are trained simultaneously in a min–max competition with respect to binary cross-entropy loss. 
The final objective for \emph{G} is to output samples that follow as close as possible the ``real'' data distribution, while \emph{D} learns to spot the fake samples from real ones, by penalizing \emph{G} for producing implausible results.



Given the success achieved in the field of image processing, GANs have also been effective in speech processing tasks. 
In this regard, WaveGAN \cite{donahue2019wavegan} represents the pioneering attempt to adapt a deep convolutional GAN (DCGAN) structure for speech, by compressing the two-dimensional image input into one-dimensional. It laid the foundations for GAN-based practical audio synthesis and for converting different image generation GANs to operate on waveforms.

Several extensions have been derived from WaveGAN; to name a few, cWaveGAN \cite{lee2018conditional}, which allows conditioning both \emph{G} and \emph{D} with additional information to drive the generation process, and Parallel WaveGAN \cite{yamamoto2020parallel}, which uses a multi-resolution STFT loss along with the adversarial loss.

As outlined in \cite{donahue2019wavegan}, in the generative setting, working with compressed time-frequency representations may be problematic as the generated spectrograms are non-invertible, hence they cannot be listened to without lossy estimations, nevertheless, the practice of bootstrapping image recognition algorithms for audio tasks has become commonplace; examples include SpecGAN \cite{donahue2019wavegan}, MelGAN \cite{kumar2019melgan}, VocGAN \cite{yang2020vocgan} and StyleGAN \cite{palkama2020conditional}.

\subsection{Pix2pix}
Pix2pix is a conditional GAN (cGAN) originally developed in 2017 by Phillip Isola, et al.~\cite{isola2017image} for synthesizing photos from label maps, reconstructing objects from edge maps and colorizing images. Unlike a vanilla GAN which uses only random noise seeds to trigger generation, a cGAN introduces a sort of supervision by feeding the generator with the target information $c$, categorical labels or contextual samples. The discriminator is also conditioned by $c$, to help distinguish more accurately the matching and alignment of two images: 
\begin{equation}
\begin{split}
    \underset{G}{\text{min }}\underset{D}{\text{max }}\mathcal{L}_{cGAN}\left( D,G\right)= &\mathbb{E}_{\textit{x,c}}\left[ \log\left( D(x|c) \right) \right] + \\
    & \mathbb{E}_{\textit{z,c}}\left[ \log\left( 1-D(G(z|c)) \right) \right]
\end{split}
\end{equation}

Unlike other cGAN-based works (e.g. \cite{mirza2014conditional}\cite{miyato2018cgans}), Isola et al. demonstrate that the input noise vector \emph{z} does not have a significant impact if the conditioning information is strong enough, so they removed it, getting the same stochastic behavior by adding dropout layers to the generator.

\section{Neural Concealment Architecture}\label{sec:proposed_method}

\begin{figure*}
    \centering
    \includegraphics[width=0.95\textwidth]{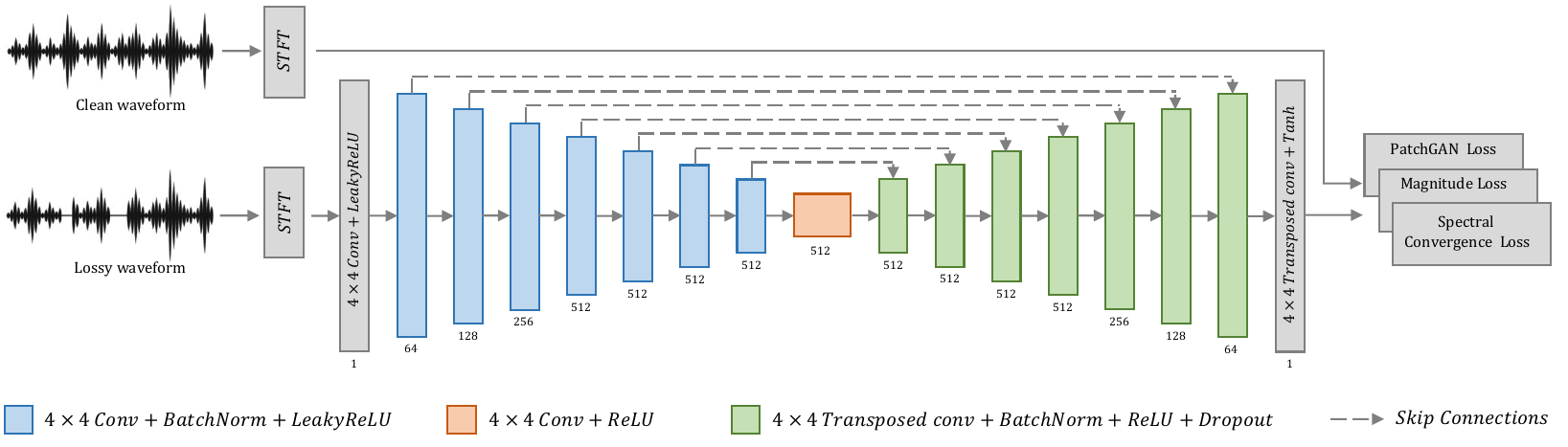}
    \caption{The proposed framework is composed of the U-Net for spectrogram inpainting. Deep feature loss for training the U-Net is obtained by ensembling the discriminator loss (binary cross-entropy between patches), along with the spectral distances ($\mathcal{L}_{mag}$ and $\mathcal{L}_{sc}$), between the representations of the recovered and the actual STFT log-magnitudes.}
    \label{fig:framework}
\end{figure*}

An overview of our bin2bin architecture is presented in Fig. \ref{fig:framework}.
The main contribution of this paper is the adaptation of the pix2pix architecture, for the audio packet loss concealment task, through an in-depth evaluation of both generative and discriminative processes, optimized to inpaint spectrograms gaps. We adopt the term bin2bin as a direct translation of pix2pix, inspired by the fundamental unit (bin) of the discretized time and frequency axes of the spectrogram.


\subsection{Generator}

In the proposed bin2bin scheme, the generator architecture makes use of the U-Net \cite{ronneberger2015u} structural design with the insertion of skip-connections between affine layers. The U-Net is composed of a convolutional encoder that down-samples the input image in the first half of the architecture, and a decoder that upsamples the latent representation applying 2D transposed-convolutions.

The clean signal $s$ and its lossy counterpart $\tilde{s}$, are first transformed into time-frequency spectrograms. In the provided implementation, all STFTs are computed with a 512 points Hann window, corresponding to 32 milliseconds at the sample rate of 16000 Hz, and a hop size of 64. The STFT parameters have been chosen to ensure a balanced resolution between the regions to be reconstructed and the reliable parts acting as conditioning contexts.

Our generator \emph{G} accepts $1 \times 256 \times 256$ inputs, where each dimension represents, respectively, the number of \emph{Channels}, \emph{Frequency} and \emph{Time} bins, hence, a portion of such size is extracted at a random time, from the aforementioned spectrograms $S$ and $\tilde{S}$, regardless of the amount of lost fragments present inside.

Only the log-magnitude spectrogram is fed into the generator; for the training stage, the phase information is discarded, while for the test stage it is used to initialize the Griffin-Lim \cite{griffin1984signal} phase reconstruction algorithm.


\subsection{Discriminator}

The discriminator is built on a custom architecture, specifically designed for the pix2pix framework, called PatchGAN \cite{isola2017image}. It is basically a fully convolutional network that maps the input image into an $N \times N$ feature map of outputs $Y$, in which each patch $y_{ij}$ indicates whether the corresponding portion of input is real or fake. The patches originate from overlapped receptive fields, which can be retrieved through simple backtracking operations. 

In the original paper \cite{isola2017image}, an ablation study was conducted to determine the best configuration of \emph{D} (number of conv. layers, kernels size), to maximize the evaluated metrics. In this work we focused on a similar aspect:
we tested the effect of varying the size of the discriminator convolutional kernels, to achieve a rectangular receptive field, instead of the square dimension ($70\times70$ pixels) used in pix2pix. We motivated this decision by observing that the portions of the spectrogram to be concealed extend over the entire frequency dimension, and a relatively small part of the time dimension. We traded-off between the complexity of \emph{D} and the desired shape, obtaining an optimal receptive field of $162\times24$, with rectangular $8 \times 2$ kernels for all conv layers. 

\subsection{Post-processing}
The generator output represents the magnitudes of the TF coefficients, both of the reliable and lost regions. The synthesis by the inverse STFT introduces an inherent cross-fading, which significantly reduces artifacts. For the phase reconstruction we used a modified version of Griffin-Lim \cite{griffin1984signal} algorithm, by providing the phase of the lossy frame as an initial estimate. In this way the synthesis of the reconstructed waveform is considerably sped up; we can obtain maximum quality, with less than 10 iterations of the algorithm.

\subsection{Loss functions}

The generator model is trained by mixing the GAN objective with a traditional pixel-wise loss, between the generated reconstruction of the source spectrogram and the expected target spectrogram.

Differently from the original paper, we have found it more beneficial to use loss functions related to the perceptual quality of the audio signal: log-STFT magnitude loss $(\mathcal{L}_{mag})$ and Spectral Convergence loss $(\mathcal{L}_{sc})$, defined as follows:

\begin{equation}
    \mathcal{L}_{mag}\left(S, \tilde{S}\right)=\frac{\sum_{t,f}\lvert \log \lvert S_{t,f} \rvert - \log \lvert \tilde{S}_{t,f} \rvert \rvert} {T \cdot N}
\end{equation}

\begin{equation}
    \mathcal{L}_{sc}\left(S, \tilde{S}\right)=\frac{\sqrt{\sum_{t,f}\left( \lvert S_{t,f}\rvert - \lvert \tilde{S}_{t,f}\rvert \right)^2}}{\sqrt{\sum_{t,f} \lvert S_{t,f} \rvert^2}}
\end{equation}
where $\lvert S_{t,f} \rvert$ and $\lvert \tilde{S}_{t,f} \rvert$ represent the STFT magnitude vector of $s$ and $\tilde{s}$ respectively, at time $t$, while $T$ and $N$ denote the number of time bins and frequency bins of a frame.

As outlined in \cite{arik2018fast}, $\mathcal{L}_{sc}$ highly emphasizes large spectral components, which helps especially in early phases of training, while $\mathcal{L}_{mag}$ accurately fits small amplitude variations, which tends to be more important towards the later phases of training.

The goal of the adversarial loss is to drive the generator model to output T-F representations that are plausible in the target domain, whereas the spectral losses regularize the generator model to output spectrograms that are a plausible translation of the source context. The combination of the adversarial loss and the spectral losses is controlled by the hyperparameters $\lambda_1$ and $\lambda_2$, both set to 250, since it has been observed that the spectral loss is more important for reconstruction than the adversarial one.

\begin{equation}
    \mathcal{L}=\mathcal{L}_{cGAN} + \lambda_1 \mathcal{L}_{mag} + \lambda_2 \mathcal{L}_{sc}
\end{equation}

The discriminator model is trained in a standalone manner in the same way as in a traditional GAN model, minimizing the negative log-likelihood of identifying real and fake images, although conditioned on the clean spectrogram, which is concatenated with $G(\tilde{S})$ to form the input of $D$.

We followed a common practice in training generative networks \cite{goodfellow2016nips}, which consists in balancing the evolution of training by iterating $n_G$ times the generator weights update, for every one of \emph{D}. We used the value $n_G=10$.


The models were trained for 50 epochs, following an early stopping policy based on the spectral losses observed on the validation set. We used the Adam \cite{kingma2014adam} optimizer with a learning rate of 0.0002 for both the generator and the discriminator, and a batch size of 8.

\section{Datasets}\label{sec:setup}
We used the VCTK Corpus (Centre for Speech Technology Voice Cloning Toolkit) \cite{yamagishi2019vctk} set of data to simulate loss traces, for training and evaluation of the speech PLC model.

\begin{figure}[htbp]
    \centerline{\includegraphics[width=0.92\columnwidth]{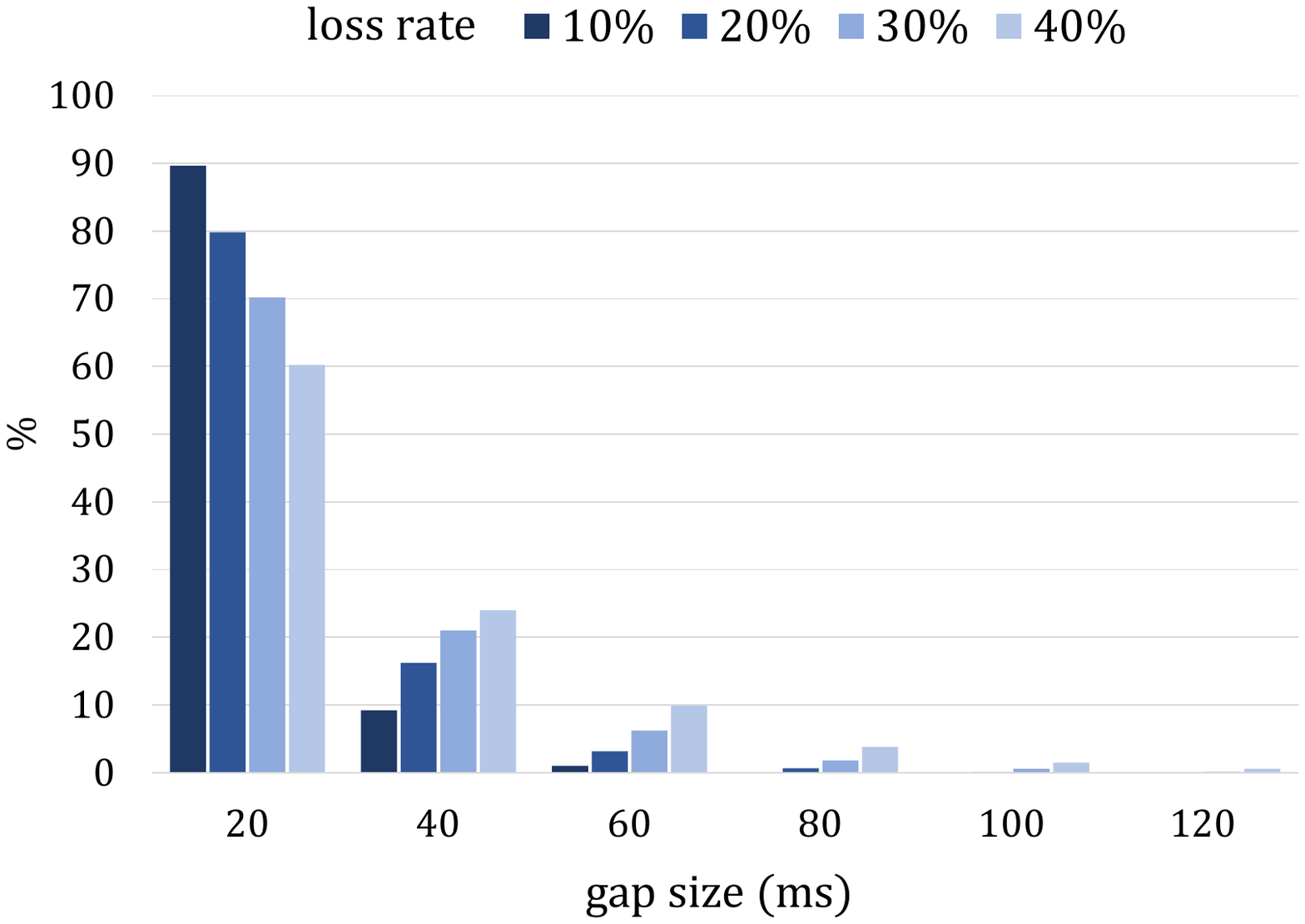}}
    \caption{Gap size distribution obtained by injecting zero-valued frames, with different rates.}
    \label{fig:data_stats}
\end{figure}

VCTK contains about 44 hours of clean speech from 109 English speakers, 47 males and 62 females, with different accents. To comply with the policy followed by the comparing methods, we downsampled the audio to 16 kHz, trimmed leading and trailing silence, and split into three subsets: train, validation and test, the latter containing 5 speakers held out from the train and validation sets.
We assumed that the lost packets have a duration multiple of 20 ms, and were simulated by zeroing samples of the clean waveform, finally we limited to 120 ms the maximum gap length, equivalent to 6 consecutive packets.
Fig. \ref{fig:data_stats} shows the distribution of lost gaps, obtained by injecting packets with rates in the range 10\% - 40\%.

\begin{figure*}[htbp]
    \centerline{\includegraphics[width=0.88\textwidth, keepaspectratio]{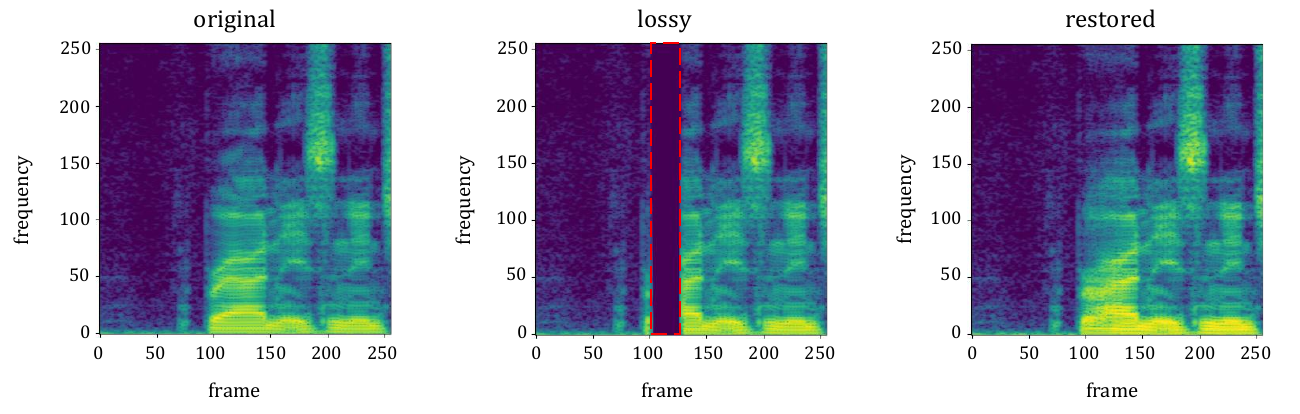}}
    \caption{Magnitude spectrograms (in dB) of an example reconstruction. Left: original signal. Center: lossy signal with 120 ms wide gap (in red-dashed box) Right: reconstruction by the bin2bin network. The axes of the plots indicate the frequency bin and the frame index.}
    \label{fig:spectrograms}
\end{figure*}

\section{Results and comparisons}\label{sec:results}
The proposed PLC method has been compared with three algorithmic solutions, represented by the general purpose codecs Opus \cite{valin2016high}, WebRTC \cite{blum2021webrtc} and Enhanced Voice Services (EVS) \cite{lecomte2015evs}, and against four state-of-the-art deep PLC methods: the wave-to-wave generative adversarial network (PLCNet) \cite{liu2022plcnet}, the mel-to-wave non-autoregressive adversarial auto-encoder (PLAAE) \cite{pascual2021adversarial}, the wave-to-wave adaptive recurrent neural network (RNN) \cite{lotfidereshgi2018speech} and the time-frequency hybrid generative adversarial network (TFGAN) \cite{wang2021temporal}. In addition, the evaluation metrics obtained by simply zero-filling the lost gaps were also reported as a baseline.

We evaluated the performances of the proposed generative inpainting method, in terms of Wide-Band Perceptual Evaluation of Speech Quality (PESQ) \cite{recommendation2001perceptual} and Short-Time Objective Intelligibility (STOI) \cite{taal2010short}. The implementations used in this paper are from \cite{miao_wang_2022_6549559} for PESQ, and from \cite{mpariente2018pystoi} for STOI. 

Table \ref{tab:pesq_stoi_1} shows the experimental results for PESQ and STOI, under different packet loss rates, compared with the PLCNet method, 
It can be seen that the proposed model can achieve a significant improvement in performance, the more the loss rate increases, so it is also able to cope better with large gaps of adjacent lost packets. The improvement is notable on PESQ scores; it ranges from +6.0\% (loss rate 10\%) to +27.5\% (loss rate 40\%). The STOI shows less noticeable gains, only for higher loss rates: +2.3\% (loss rate 30\%) and +7.8\% (loss rate 40\%).

Table \ref{tab:pesq_stoi_2} summarizes the results of the proposed method with all the competing approaches. Values represent the average score of PESQ and STOI under all packet loss rates investigated.
Compared with the best performing network among previous state-of-the-art systems (PLCNet), bin2bin improves PESQ by 15.3\% and STOI by 2.4\%, while, in comparison with the best codec-based concealment (EVS), the improvement rises up to 43.9\% for PESQ and 12.8\% for STOI.

Figure \ref{fig:spectrograms} shows the qualitative results of a concealed 120 ms wide gap, within a test sample. This represents the worst scenario, in terms of extent of lost fragments, the network is trained to face.

In addition, we timed the forward execution of the bin2bin inpainting process, both in a CPU environment (Intel core i7-6850K) and a GPU environment (Nvidia Titan Xp), obtaining real-time (RT) factor values of 0.17 and 0.11 respectively.

\begin{table}[htbp]
\centering
\caption{Objective scores for bin2bin and PLCNet, under different packet loss rates}
\label{tab:pesq_stoi_1}
\footnotesize
    \begin{tabular}{@{\extracolsep{8pt}}ccccc}
    \midrule
      & Packet Loss Rate & zero-fill & PLCNet & bin2bin\\
    \midrule
    \multirow{4}{*}{PESQ} & 10 \% & 2.13 & 3.12 & \bf{3.31}\\
                          & 20 \% & 1.04 & 2.60 & \bf{2.87}\\
                          & 30 \% & 0.89 & 2.04 & \bf{2.50}\\
                          & 40 \% & 0.81 & 1.71 & \bf{2.18}\\
    \midrule
    \multirow{4}{*}{STOI} & 10 \% & 0.86 & \bf{0.93} & 0.92\\
                          & 20 \% & 0.81 & \bf{0.90} & \bf{0.90}\\
                          & 30 \% & 0.73 & 0.85 & \bf{0.87}\\
                          & 40 \% & 0.61 & 0.77 & \bf{0.83}\\
    \bottomrule
\end{tabular}
\end{table}


\begin{table}[htbp]
\centering
\caption{Average objective scores for the comparison PLC solutions, under packet loss rate 10\%-40\%}
\label{tab:pesq_stoi_2}
\footnotesize
    \begin{tabular}{@{\extracolsep{8pt}}ccc}
    \midrule
    & PESQ & STOI\\
    \midrule
    bin2bin & \bf{2.72} & \bf{0.88}\\
    PLCNet & 2.36 & 0.86\\
    PLAAE & 2.04 & 0.84\\
    TF-GAN & 1.97 & 0.81\\
    RNN & 1.91 & 0.77\\
    EVS & 1.89 & 0.78\\
    Opus & 1.77 & 0.77\\
    WebRTC & 1.70 & 0.70\\
    zero-fill & 1.22 & 0.75\\
    \bottomrule
\end{tabular}
\end{table}

\section{Conclusions}\label{sec:conclusions}

In this paper, we proposed an end-to-end pipeline for spectrogram inpainting and audio concealment using a cGAN-based architecture, inspired by the popular pix2pix framework. We combined the classical discriminative loss with a linear combination of two loss functions, that are correlated with the perceptual quality of speech. In addition, we adapted the receptive field of the PatchGAN discriminator and we used a custom initialization of the Griffin-Lim algorithm to speed up post-processing.
We demonstrated experimentally that the proposed method is capable of simultaneously identifying and recovering missing parts, thus outperforming the state-of-the-art DNN method by +15.3\% on PESQ and +2.4\% on STOI, respectively.
Finally, inference time evaluation suggests that this approach can be integrated into a real-time application, even with a mid-range hardware setting.

As future developments we plan to investigate the generator to directly process complex-valued spectrograms, in order to incorporate the phase reconstruction directly into the generative model.

\setstretch{0.97}
\bibliography{biblio.bib}{}
\bibliographystyle{IEEEtran}

\end{document}